\documentclass{article}

\usepackage[nonatbib,final]{nips_2017}

\usepackage[utf8]{inputenc} 
\usepackage[T1]{fontenc}    
\usepackage{hyperref}       
\usepackage{url}            
\usepackage{booktabs}       
\usepackage{amsfonts}       
\usepackage{amssymb}
\usepackage{amsmath}
\usepackage{bbm}
\usepackage{nicefrac}       
\usepackage{microtype}      
\usepackage{caption}
\usepackage{float}
\usepackage{graphicx}
\usepackage{wrapfig}
\usepackage{authblk}
\usepackage{subcaption}

\graphicspath{ {./} }

\title{A multi-modal neural network for learning cis and trans regulation of stress response in yeast}

\author[1,2,3]{Boxiang Liu}
\author[4]{Nadine Hussami}
\author[3,5]{Avanti Shrikumar}
\author[3]{Tyler Shimko}
\author[6]{Salil Bhate}
\author[6]{Scott Longwell}
\author[2,3]{Stephen Montgomery}
\author[3,6]{Anshul Kundaje}
\affil[ ]{$^1$Departments of Biology, $^2$Pathology, $^3$Genetics, $^4$Statistics, $^5$Computer Science, and $^6$Bioengineering, Stanford University}
\affil[ ]{\textit{\{bliu2,nadinehu,avanti,tshimko,bhate,longwell,smontgom,akundaje\}@stanford.edu}}

\begin{document}

\maketitle

\begin{abstract}

\small Deciphering gene regulatory networks is a central problem in computational biology. Here, we explore the use of multi-modal neural networks to learn predictive models of gene expression that include cis and trans regulatory components. We learn models of stress response in the budding yeast \textit{Saccharomyces cerevisiae}. Our models achieve high performance and substantially outperform other state-of-the-art methods such as boosting algorithms that use pre-defined cis-regulatory features. Our model learns several cis and trans regulators including well-known master stress response regulators. We use our models to perform in-silico TF knock-out experiments and demonstrate that in-silico predictions of target gene changes correlate with the results of the corresponding TF knockout microarray experiment.
  
\end{abstract}

\section{Introduction}

Gene transcription is regulated by transcription factor (TFs) complexes that bind specific sequence motifs encoded in the DNA of cis-regulatory elements. Learning predictive models of gene expression that integrate cis and trans regulatory information is a critical first step to decipher the effects of natural and disease-associated perturbations to regulatory networks. Most approaches for learning transcriptional regulatory models typically use compendia of known TF binding sequence motifs to represent the DNA sequence of cis-regulatory elements \cite{Bussemaker:2001ft,Phuong:2004kk,Soinov:2003iz,Segal:2003ks,Middendorf:2004gta,Ruan:2006hl}. However, these TF motif compendia are often incomplete and consist of position weight matrices (PWMs) which may not be optimal representations for predictive models. Recently, convolutional neural networks have been used to learn de-novo representations from raw regulatory DNA sequence that can predict TF binding, chromatin accessibility as well the effects of non-coding variants on these molecular phenotypes via in-silico mutagenesis\cite{Kelley:2016bv}. However, these cis-regulatory deep learning models do not model the effects of trans factors and are hence incapable of predicting gene expression in different cellular states.

Here, we present a multi-modal deep neural network architecture that can be used to predict gene expression of any target gene in any cellular state based on the raw cis-regulatory sequence of the gene and the expression levels of trans factors (TFs and signaling molecules) in that cellular state. We learn predictive regulatory programs of stress response in the budding yeast \textit{Saccharomyces cerevisiae}\cite{Gasch:2000wl}. As compared to mammals, yeast has a relatively simple cis-regulatory architecture governed primarily by the promoter sequence directly upstream of genes; has a far more comprehensive set of known trans factors (TFs and signaling molecules) and TF binding sequence motifs and has an extensive collection of perturbation experiments such as TF knockouts. Yeast is hence an ideal model organism to systematically compare the performance of our deep learning approach to other alternatives that use engineered features and to benchmark the ability of these models to predict the effects of perturbations such as TF knockouts.

\section{Methodology}
We formulate the problem as a supervised learning task, where the goal is learn a model $E_{g,c} = F(S_g , T_c)$ that can predict gene expression ($E_{g,c}$) of any gene ($g$) in any stress condition ($c$) based on two complementary regulatory inputs - a cis component represented by the raw promoter sequence ($S_g$) of the gene $g$ and a trans component represented by the expression of all trans factors ($T_c$) in condition $c$. We use a multimodal neural network architecture  that includes a convolutional sequence module to learn predictive patterns from raw promoter sequences, a dense module to derive features from regulator expression and an integration module that learns cis-trans interactions.

\subsection{Cis regulatory sequence module}
The input to this module is the raw 1Kb promoter sequence ($S_g$) of a gene ($g$) which is represented using a standard 4 channel (A,C,G,T) one-hot encoding. We use two convolutional layers each of which contain 50 filters (size 9, stride 1) with ReLU activations. These layers are followed by a max pooling layer (size 4, stride 4). We apply batch normalization before each ReLU activation to mitigate covariate shifts during training and to accelerate learning. We train on forward and reverse complements of each sequence and also use reverse complement aware convolutional filters via parameter sharing \cite{Shrikumar103663}. The final layer of this module is a dense layer with 512 units.

\subsection{Trans regulator expression module}
The input to this module is a vector of expression levels ($T_c$) of 472 known transcription factors and signal molecules (kinases, phosphatases, receptors) in stress condition $C$. The input feeds into a dense layer of 512 hidden units.

\subsection{Integration module}
We concatenate the outputs of the cis and trans module and use 2 dense layers with 512 units integrate the cis and trans modules. The final layer feeds into a linear neuron (for regression) or a softmax neuron (for multi-class classification). 

\subsection{Gene expression data}
The gene expression dataset is a microarray dataset\cite{Gasch:2000wl} that spans 6100 genes under 173 diverse stress conditions. The measurements were given as $log_2$ expression values representing the fold change w.r.t. the untreated reference condition. For the regression models, the expression levels were used as is. For the classification models, we discretized expression into 3 classes in $\{-1,0,1\}$ to represent upregulation, baseline, and downregulation, such that expression values in $[-\infty,-0.5]$ are converted to -1, $[-0.5,0.5]$ to 0, and $[0.5,\infty]$ to +1. We used a 80-10-10 split of $(g,c)$ spanning all genes and all stress conditions to create training, validation, and test datasets.

\subsection{Training methodology}
We optimize square loss (for regression models) or softmax loss (for classification models) using SGD with a learning rate of 0.01 and momentum of 0.5. We decrease the learning rates by half every five epoch if no improvement in validation accuracy is observed. All experiments were performed on Nvidia GeForce GTX 970 using Keras 1.2.2 with Tensorflow backend.

\section{Results}

\subsection{Prediction performance}
\begin{figure}[h]
    \centering
    \begin{subfigure}{.475\textwidth}
        \includegraphics[width=0.9\textwidth]{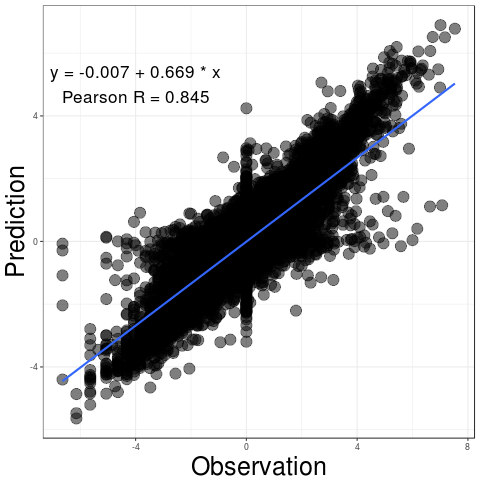}
        \caption[Correlation]{Correlation}
        \label{fig:correlation}
    \end{subfigure}\hfill
    \parbox{.475\textwidth}{
        \begin{subfigure}{.475\linewidth}
            \includegraphics[width=\textwidth]{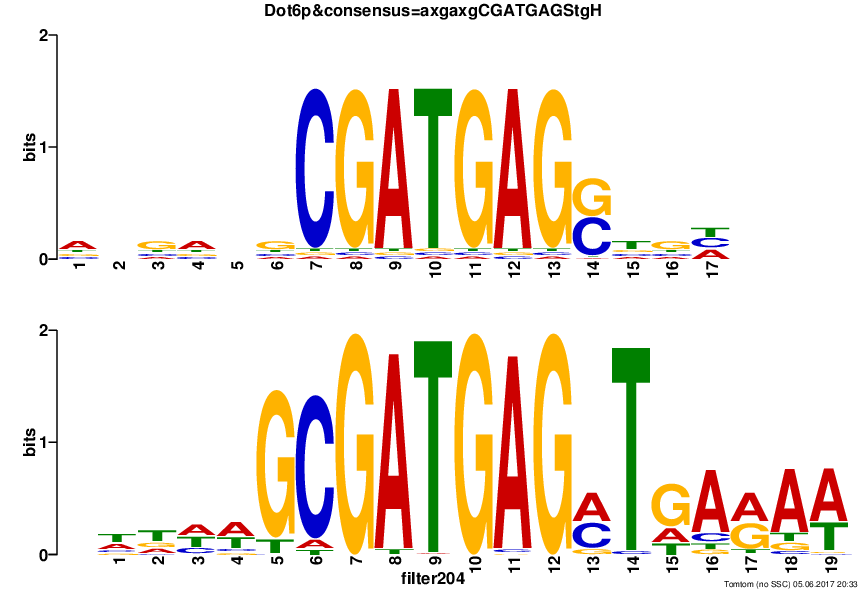}
            \caption{DOT6P}
    \end{subfigure}
    \begin{subfigure}{.475\linewidth}
            \includegraphics[width=\textwidth]{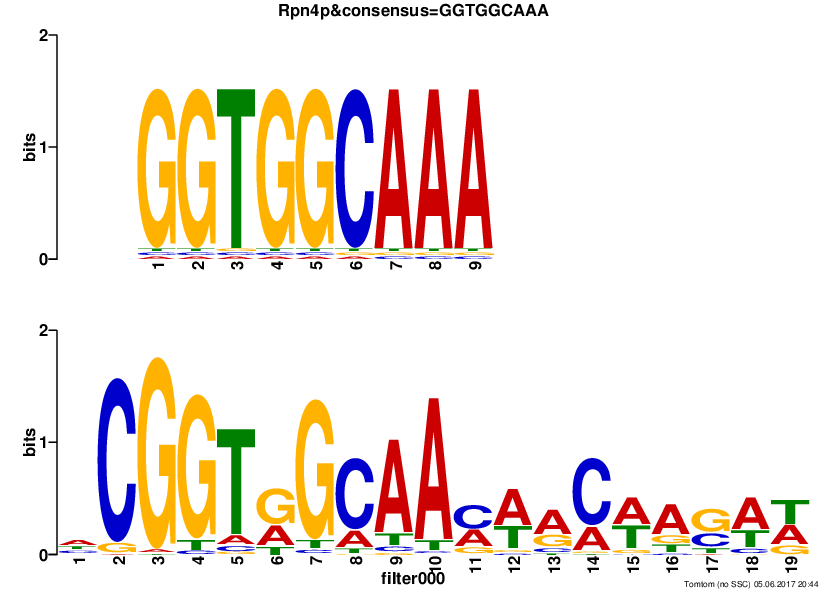}
            \caption{RPN4P}
    \end{subfigure}\\
    \begin{subfigure}{.475\linewidth}
            \includegraphics[width=\textwidth]{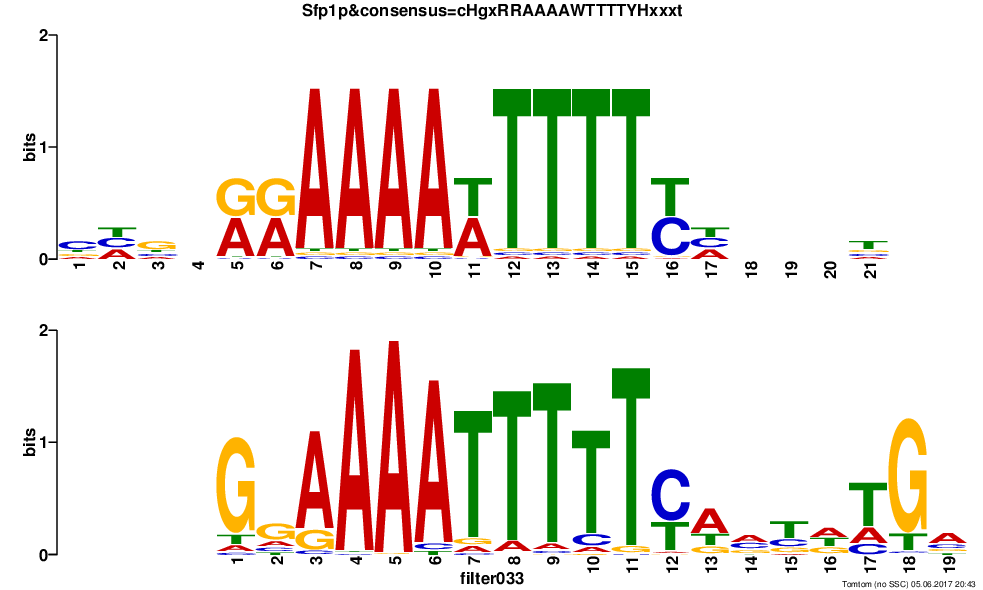}
            \caption{SFP1P}
    \end{subfigure}
    \begin{subfigure}{.475\linewidth}
            \includegraphics[width=\textwidth]{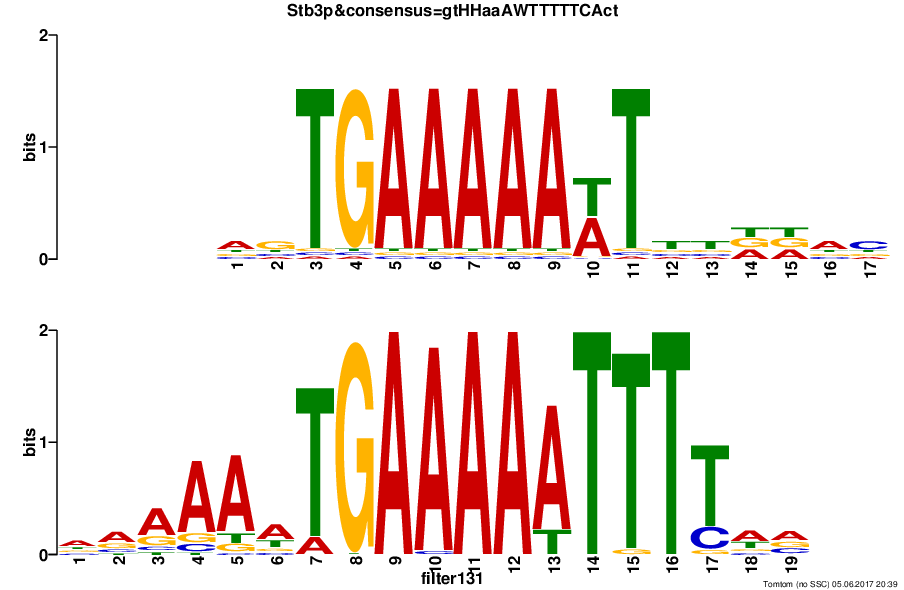}
            \caption{STB3P}
    \end{subfigure}
    }
    \caption{(a) Predicted vs grounth truth (b-e) examples of recovered motifs}\label{fig:1}
\end{figure}

We compared our model against two state-of-the-art models, GeneClass \cite{Middendorf:2004gta} and BDTree \cite{Ruan:2006hl} that use the multi-class classification formulation on the same dataset \cite{Gasch:2000wl} but represent the cis-regulatory sequence as a vector of motif occurrences spanning all known yeast motifs. The GeneClass model is a boosted alternating decision tree, and the BDTree model is a bidirectional regression tree. For the classification tasks, our best model outperformed the previous state-of-the-art by 16.6\% (Table \ref{table:classification performance}). Hence, the neural network model that learns de-novo representations from raw sequence significantly outperform ensemble models trained using known motifs. In addition, our model achieved a high Pearson correlation of 0.845 for the regression task on the test set (Figure \ref{fig:correlation}).

\begin{table}[!hbt]
\caption{Classification performance}
\centering
 \begin{tabular}{|c | c|} 
 \hline
 Method & Accuracy \\
 \hline
 GeneClass & 60.9\% \\ 
 \hline
 BDTree & 62.9\% \\
 \hline
 DNN & \textbf{79.5\%} \\
 \hline
\end{tabular}
\quad
\centering
 \begin{tabular}{|c | c c c|} 
 \hline
 {} & \multicolumn{3}{|l|}{Predicted by DNN} \\
  & Down & Baseline & Up \\ 
 \hline
 Down & 10.14 & 7.47 & 0.13 \\
 \hline
 Baseline & 3.29 & 59.77 & 3.02 \\
 \hline
 Up & 0.18 & 6.42 & 9.59 \\
 \hline
\end{tabular}
\label{table:classification performance}
\end{table}

\subsection{Deciphering predictive cis-regulatory sequence features}
Unlike GeneClass and BDTree which rely on existing motif annotations, our model has the potential to learn known and novel sequence features directly from the raw promoter sequences. We used a method similar to Basset \cite{Kelley:2016bv} to identify sequence patterns that activate the convolutional filters. For each convolutional filter, we select the 100 sequence segments with the highest activation. We next calculate the PWM based on nucleotide frequency in these sequence segments. To test whether any PWM correspond to known motif, we used TomTom (http://meme-suite.org/tools/tomtom) to compare against the YEASTRACT database. We found that our model learns both known (Figure 1) and \textit{de novo} motifs.

\subsection{Deciphering predictive trans regulators of stress response}
We estimated the importance of each trans-regulator for each training/test example as the gradient (w.r.t to output) times the magnitude of input (G-by-I). We summed the G-by-I values across all genes and conditions to obtain the global estimate of trans-regulator importance. The top ranking trans-regulators include several well-known regulators of stress response in yeast such MSN2/4 (master stress response TF), TPK1 (kinase that phosphorylates MSN2/4 which controls its cellular localization), USV1, PPT1, XBP1 and YVH1.

\begin{table}[!hbt]
\caption{Rank of regulator module inputs}
\centering
 \begin{tabular}{|c | c|} 
 \hline
 Rank & Regulator \\
 \hline
 1 &  USV1 / YPL230W \\ 
 \hline
 2 & DAL80 / YKR034W \\
 \hline
 3 & XBP1 / YIL101C \\
 \hline
 4 & PPT1 / YGR123C \\
 \hline
 5 & LSG1 / YGL099W \\
 \hline
 6 & CIN5 / YOR028C \\
 \hline
 7 & YVH1 / YIR026C \\
 \hline
 8 & TPK1 / YJL164C \\
 \hline
 9 & GAC1 / YOR178C \\
 \hline
 10 & MSN4 / YKL062W \\
 \hline
\end{tabular}
\label{table:rank of regulator module inputs}
\end{table}

\subsection{Predicting the effects of a transcription factor knockout}
\begin{figure}
\centering
\begin{subfigure}{.5\textwidth}
  \centering
  \includegraphics[width=0.8\textwidth]{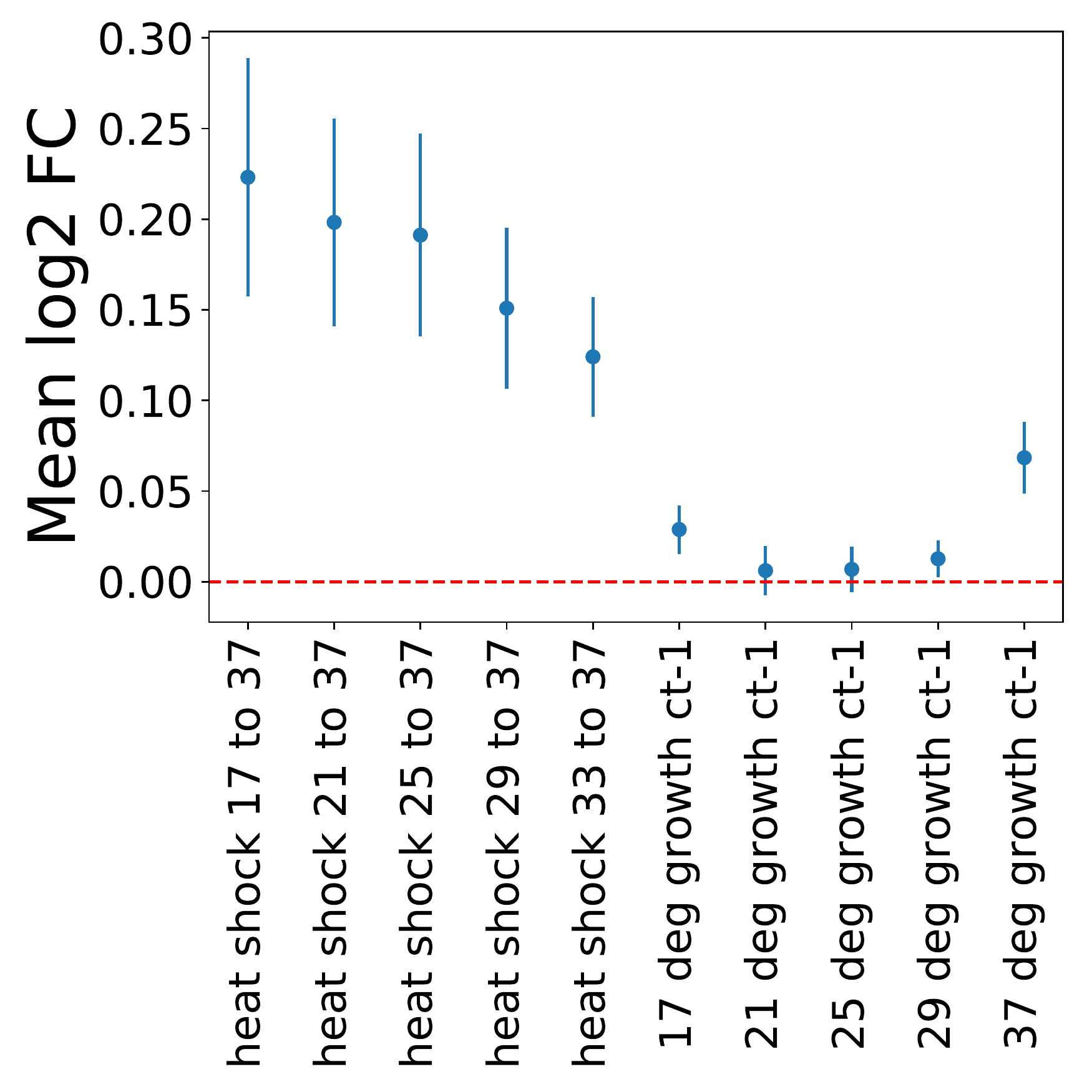}
  \caption{Target gene expression change}
  \label{fig:target gene expression change}
\end{subfigure}%
\begin{subfigure}{.5\textwidth}
  \centering
  \includegraphics[width=1.0\textwidth]{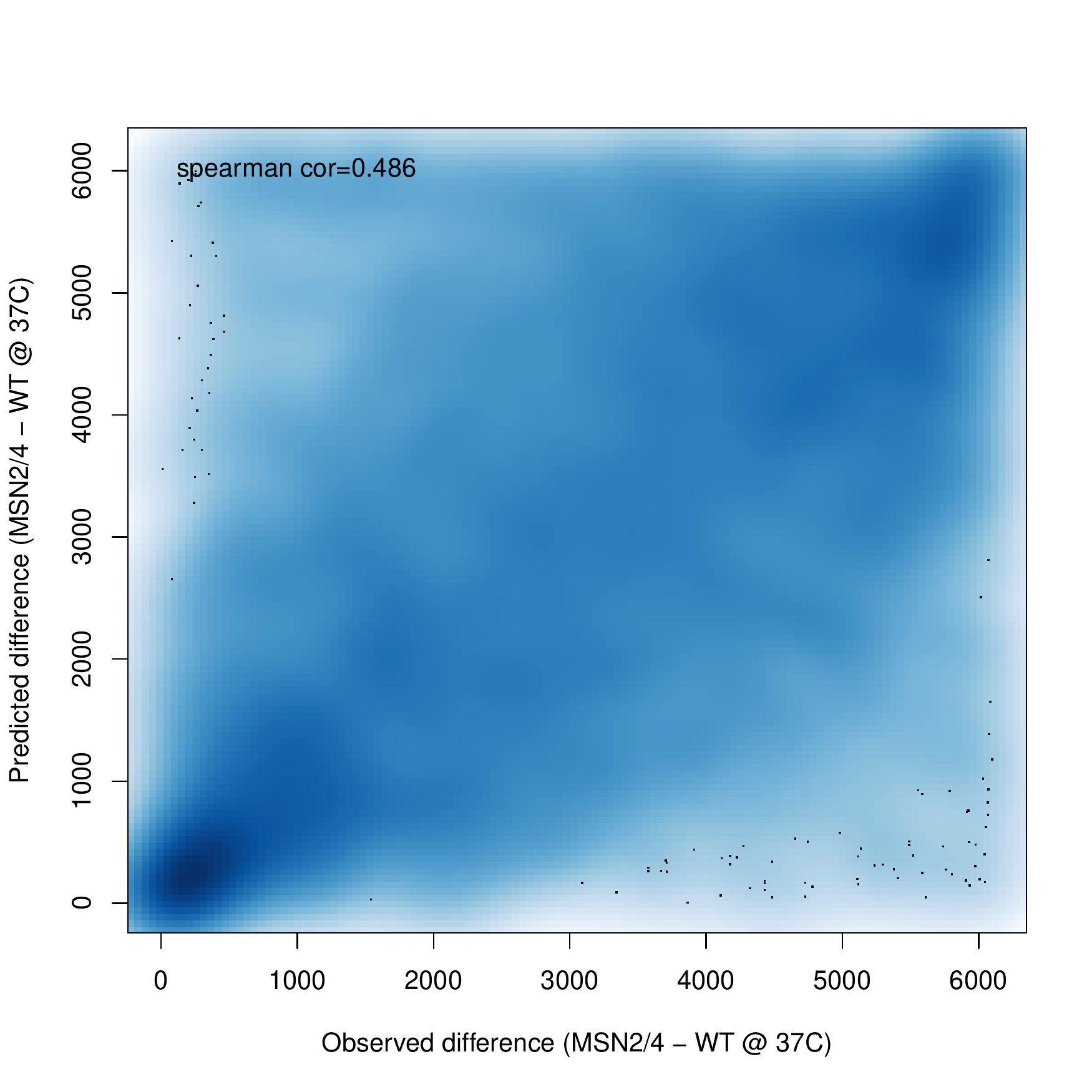}
  \caption{Predicted vs ground truth}
  \label{fig:predicted vs ground truth msn24 ko}
\end{subfigure}
\caption{(a) The expression of MSN2/4 known target gene experience larger change under stress conditions. (b) Predicted expression change vs actual microarray experiment.}
\label{fig:fig2}
\end{figure}

The ultimate test of a predictive model of gene expression is based on its ability to predict the effect of some previously unseen perturbation. Therefore, we performed \textit{in silico} knockout experiment on MSN2/4, where we replaced all instances of its motif 'AGGGG' with neutral 'NNNNN', and reduced the expression level of MSN2/4 by 32 fold. MSN2/4 are activated under heat shock but are inactive under steady-state growth \cite{Sadeh:2011eg}. As expected, we observed that MSN2/4 target genes experience greater change under heat shock conditions than steady-state growth condition (Fig. \ref{fig:target gene expression change}). We also compared our prediction against microarray measurement in a actual MSN2/4 knockout strain, and observed significant correlation between the two (spearman correlation = 0.486, Fig. \ref{fig:predicted vs ground truth msn24 ko}). 

\section{Discussion}
We present a multi-modal deep learning architecture that can accurately predict the gene expression response of yeast to various stress conditions as a function of cis-regulatory sequence and trans factor expression. Our model outperforms other approaches that use engineered features. Preliminary analyses of the globally predictive features indicate that our models captures several well known stress response factors. We are currently exploring the context-specific cis and trans regulators learned by the model at the level of individual genes and gene modules across the diversity of stresses. Further, through \textit{in silico} knockout of a key stress response factor, we show that our model makes reasonably accurate predictions similar to the true knockout microarray experiment. It is worth noting that while the true knockout experiment measures the direct and indirect effects of the TF knockout, our model is more likely to predict direct effects. We are using in-vivo TF binding maps of the TF to distinguish the direct from indirect targets of MSN2/4 to further investigate this issue. This work lays the foundation for developing novel neural network architectures to model transcriptional regulation in mammalian systems.

\bibliographystyle{unsrt}
\bibliography{main}

\begin{thebibliography}{10}

\bibitem{Bussemaker:2001ft}
Harmen~J Bussemaker, Hao Li, and Eric~D Siggia.
\newblock {Regulatory element detection using correlation with expression}.
\newblock {\em Nature Genetics}, 27(2):167--174, February 2001.

\bibitem{Phuong:2004kk}
Tu~Minh Phuong, Doheon Lee, and Kwang~Hyung Lee.
\newblock {Regression trees for regulatory element identification}.
\newblock {\em Bioinformatics}, 20(5):750--757, March 2004.

\bibitem{Soinov:2003iz}
Lev~A Soinov, Maria~A Krestyaninova, and Alvis Brazma.
\newblock {Towards reconstruction of gene networks from expression data by
  supervised learning}.
\newblock {\em Genome biology}, 4(1):R6, January 2003.

\bibitem{Segal:2003ks}
Eran Segal, Michael Shapira, Aviv Regev, Dana Pe'er, David Botstein, Daphne
  Koller, and Nir Friedman.
\newblock {Module networks: identifying regulatory modules and their
  condition-specific regulators from gene expression data}.
\newblock {\em Nature Genetics}, 34(2):166--176, June 2003.

\bibitem{Middendorf:2004gta}
Manuel Middendorf, Anshul Kundaje, Chris Wiggins, Yoav Freund, and Christina
  Leslie.
\newblock {Predicting genetic regulatory response using classification}.
\newblock {\em Bioinformatics}, 20(suppl 1):i232--i240, August 2004.

\bibitem{Ruan:2006hl}
Jianhua Ruan and Weixiong Zhang.
\newblock {A bi-dimensional regression tree approach to the modeling of gene
  expression regulation.}
\newblock {\em Bioinformatics}, 22(3):332--340, February 2006.

\bibitem{Kelley:2016bv}
David~R Kelley, Jasper Snoek, and John~L Rinn.
\newblock {Basset: learning the regulatory code of the accessible genome with
  deep convolutional neural networks.}
\newblock {\em Genome research}, 26(7):990--999, July 2016.

\bibitem{Gasch:2000wl}
A~P Gasch, P~T Spellman, C~M Kao, O~Carmel-Harel, M~B Eisen, G~Storz,
  D~Botstein, and P~O Brown.
\newblock {Genomic expression programs in the response of yeast cells to
  environmental changes.}
\newblock {\em Molecular Biology of the Cell}, 11(12):4241--4257, December
  2000.

\bibitem{Shrikumar103663}
Avanti Shrikumar, Peyton Greenside, and Anshul Kundaje.
\newblock Reverse-complement parameter sharing improves deep learning models
  for genomics.
\newblock {\em bioRxiv}, 2017.

\bibitem{Sadeh:2011eg}
A~Sadeh, N~Movshovich, M~Volokh, L~Gheber, and A~Aharoni.
\newblock {Fine-tuning of the Msn2/4-mediated yeast stress responses as
  revealed by systematic deletion of Msn2/4 partners. - PubMed - NCBI}.
\newblock {\em Molecular Biology of the Cell}, 22(17):3127--3138, August 2011.

\end{thebibliography}

\end{document}